\documentclass[%
 reprint,
 amsmath,amssymb,
 aps,
]{revtex4-2}

\usepackage{graphicx}
\usepackage{epstopdf}
\usepackage{dcolumn}
\usepackage{bm}
\usepackage{color}
\usepackage[mathlines]{lineno}
\usepackage[breaklinks=true,colorlinks=true,linkcolor=blue,urlcolor=blue,citecolor=blue]{hyperref}
\usepackage{ragged2e}
\usepackage{enumitem}
\usepackage{bibunits}
\usepackage{amsthm}
\usepackage{diagbox}

\begin{document}

\preprint{APS/123-QED}

\title{Interactive diversity disrupts cyclic dominance but maintains cooperation in spatial social dilemma games}

\author{Danyang Jia$^1$}
\author{Chen Shen$^{2}$}
\author{Xiangfeng Dai$^{3,4}$}
\author{Junliang Xing$^{1}$}
\email{jlxing@tsinghua.edu.cn}
\author{Pin Tao$^{1}$}
\author{Yuanchun Shi$^{1}$}
\author{Zhen Wang$^4$}
\email{zhenwang0@gmail.com}
\affiliation{
\vspace{2mm}
\mbox{1. Department of Computer Science and Technology, Tsinghua University, Beijing, 100084 China}
\mbox{2. Faculty of Engineering Sciences, Kyushu University, Fukuoka, 816-8580, Japan}
\mbox{3. School of Mechanical Engineering, Northwestern Polytechnical University, Xi'an, 710072 China}
\mbox{4. School of Artificial Intelligence, OPtics and ElectroNics (iOPEN), Northwestern Polytechnical University, Xi'an, 710072 China}
}

\date{\today}

\begin{abstract}
Cyclic dominance has emerged as a key factor in upholding cooperation within structured populations. However, this understanding has primarily been based on node dynamics, where players are constrained to employ the same strategy with all neighbors. Overlooked is the impact of interactive diversity--where players can adapt different responses to distinct neighbors--on cyclic dominance dynamics and the broader cooperation patterns. This study explores the overlooked role of interactive diversity in cyclic dominance and cooperation using a voluntary prisoner's dilemma model on various network structures. We distinguish between `node players', who use a consistent strategy with all neighbors, and `link players', who adapt their strategies based on specific interactions, influenced by both direct and indirect emotions. While direct emotion dictates the strategy between two interacting players, indirect emotion encompasses third-party influences on strategic decisions. Monte Carlo simulations uncover a multifaceted relationship: interactive diversity generally disrupts cyclic dominance, yet it can either amplify or diminish cooperation based on the inclination towards indirect strategy formulation. This suggests that the importance of cyclic dominance in underpinning cooperation might have been overstated, given that cooperation can endure even when cyclic dominance wanes due to interactive diversity.

\begin{description}
\item[Key words]Interactive diversity, Cyclic dominance, Direct delivery,Upstream delivery.
\end{description}
\end{abstract}

\maketitle


\section{Introduction}
Cyclic dominance, characterized by a set of at least three strategies where the first dominates the second, the second dominates the third, and the third in turn dominates the first, has been demonstrated to wield a pivotal role across various ecological scenarios~\cite{szabo2004rock,sasaki2007probabilistic,szolnoki2009phase,szolnoki2014pairwise}. These encompass intricate dynamics such as predator-prey interactions~\cite{cazaubiel2017collective}, the mating tactics of side-blotched lizards~\cite{sinervo1996rock}, biodiversity preservation~\cite{reichenbach2007mobility}, competition dynamics among bacterial strains~\cite{durrett1997allelopathy}, and the maintenance of cooperation within social dilemmas~\cite{szolnoki2009phase,szolnoki2014cyclic}. Among the most striking phenomena in ecology, behavioral science, economics, and other systems is the evolution of cooperation among individuals with unrelated individuals~\cite{szabo2002phase}.

The intricate comprehension of the driving mechanisms behind the evolution of cooperation remains a formidable challenge. This challenge arises from the apparent paradox that helping others can potentially undermine an individual's personal gains--a contradiction seemingly at odds with the fundamental ``survival of the fittest" principle. According to this principle, defectors who contribute nothing should logically outcompete cooperators and come to dominate the entire population. To tackle this puzzle, evolutionary game theory provides a general framework for investigating the emergence and persistence of cooperation~\cite{maynard1976evolution,weibull1997evolutionary}. Within this framework, cyclic dominance has been recognized in various systems that uphold cooperation. Notably, these systems often require several additional social mechanisms beyond mere cooperation and defection. These encompass voluntary participation (also known as ``loners")~\cite{hauert2002volunteering,hauert2002replicator,szabo2002phase}, prosocial exclusion~\cite{sasaki2013evolution,li2015social}, jokers~\cite{arenas2011joker,requejo2012stability}, peer~\cite{hauert2002volunteering,helbing2010evolutionary} and pool~\cite{sigmund2010social,szolnoki2011phase} punishment, rewards~\cite{szolnoki2010reward,sigmund2001reward}, hedgers~\cite{guo2020novel}, and exiters~\cite{shen2021exit,shen2023exit}, among others~\cite{szolnoki2014cyclic}. Importantly, the role of population structures in triggering cyclic dominance via these social mechanisms cannot be overlooked. 

These population structures primarily involve well-mixed populations and structured populations. The former assumes global interactions where players can engage with any other player in the population. Conversely, the latter assumes local interactions, restricting players to interactions with their neighboring individuals along established links. Despite this subtle distinction, cyclic dominance is more prevalent in structured populations compared to well-mixed ones. For instance, in the extended public goods game incorporating prosocial and antisocial peer punishment~\cite{szolnoki2017second}, cyclic dominance resulting from peer punishment is observed exclusively within networked populations. In contrast, well-mixed populations struggle to sustain cooperation~\cite{rand2010anti,rand2011evolution}. A similar pattern emerges in the context of the compulsory public goods game featuring prosocial pool punishment~\cite{szolnoki2011phase}, where structured populations exhibit cyclic dominance involving cooperation, defection, and prosocial pool punishment, while well-mixed populations lead to defection as the sole stable state~\cite{sigmund2010social}. Furthermore, in the context of the prisoner's dilemma game involving the loner strategy~\cite{szabo2002evolutionary}, although a cyclic dominance pattern can be observed among the strategy pairs (where defection dominates cooperation, which dominates loners, which, in turn, dominates defection), the transition of loners towards cooperation necessitates a specific degree of cooperative behavior. This ultimately leads to the dominance of loners in the optional prisoner's dilemma game~\cite{hauert2002replicator}. Networked populations ensure a large-scale cyclic dominance phenomenon, thereby facilitating the emergence of cooperation. Several other examples underscore the critical role of population structures in shaping the cyclic dominance phenomenon~\cite{masuda2006networks,frean2001rock}. In essence, these distinctions highlight the pivotal role of spatial structures in shaping cyclic dominance—a role that remains concealed for some social mechanisms within well-mixed populations.

It's noteworthy that many existing studies on cyclic dominance in structured populations assume individuals follow node dynamics~\cite{szolnoki2014cyclic}, interacting simultaneously with all partners using the same strategy, simplifying decision-making. However, while node dynamics might suit simple life forms, it fails to capture the behavior of more complex organisms in reality. In actual interactions, individuals generally follow link dynamics—an interactive diversity framework—allowing them to employ different strategies with different neighbors~\cite{wardil2009adoption,wardil2010distinguishing}. For instance, they might reciprocate cooperation with cooperative neighbors and retaliate against defective ones. Although research on link dynamics remains limited, it demonstrates that this basic mechanism yields intricate dynamics in the evolution of cooperation~\cite{sendina2020diverse}. Scholars within the evolutionary game theory framework have highlighted the pivotal role of interactive diversity in promoting cooperation within two-strategy games~\cite{su2017evolutionary,su2018evolution,jia2020evolutionary}, considering factors such as network topology, population structure, and strategy learning patterns (imitation-driven or exploration-driven). Nevertheless, the influence of interactive diversity on cyclic dominance and the subsequent evolution of cooperative behavior remains unexplored.

To address this question, we employ the voluntary prisoner's dilemma game~\cite{szabo2002evolutionary}, where players can choose between cooperation for the benefit of others, defection for self-interest, and being a loner for risk aversion. This model serves as a metaphor to investigate the impact of interactive diversity on cyclic dominance dynamics. In this study, we specifically examine the influence of interactive diversity within three types of homogeneous networks: regular lattices, regular small-world networks, and regular random graphs. The distinction in cyclic dominance across these networks lies in regular lattices facilitating local cyclic dominance, while the latter two can lead to global cyclic dominance. To elaborate on our approach, we categorize the population into two groups: node players and link players. Node players can only adopt the same strategy for different neighbors and update their strategies by imitating one of their neighbors. In contrast, link players have the flexibility to choose suitable strategies for various neighbors and update each of their strategies towards different neighbors with a certain probability, influenced by their direct and indirect emotions. Direct emotion is expressed as ``my decision towards you depends on your performance towards me"~\cite{roberts2008evolution,imhof2010stochastic,schmid2022direct}, driving the strategy to be passed between two individuals (direct delivery). Indirect emotion is expressed as ``the actions of third-party individuals (e.g., neighbor Cindy) towards me (e.g., Bob) influence my decision towards you (e.g., neighbor Alice)"~\cite{nowak2007upstream,iwagami2010upstream,herne2013experimental}, leading to the strategy being passed among unrelated individuals (upstream delivery). 

Through extensive Monte Carlo simulations, we observe that, on the one hand, interactive diversity impedes the establishment of cyclic dominance, regardless of the link players' strategy delivery methods, with this detrimental effect becoming more pronounced as the density of link players increases. On the other hand, taking interactive diversity into account leads to non-trivial results concerning the evolution of cooperation. In the case of a weak preference for upstream delivery, interactive diversity significantly enhances cooperative behavior, whereas cooperative behavior weakens under strong preferences for upstream delivery. Taken together, our results suggest that cyclic dominance, while traditionally considered a critical driver of cooperation, may be less influential in networked populations when we consider the more realistic factor of interactive diversity. This is because, in such scenarios, interactive diversity has the potential to eliminate cyclic dominance, yet cooperation can still persist and thrive.

\begin{figure*}
\centering
  \includegraphics[scale=0.55]{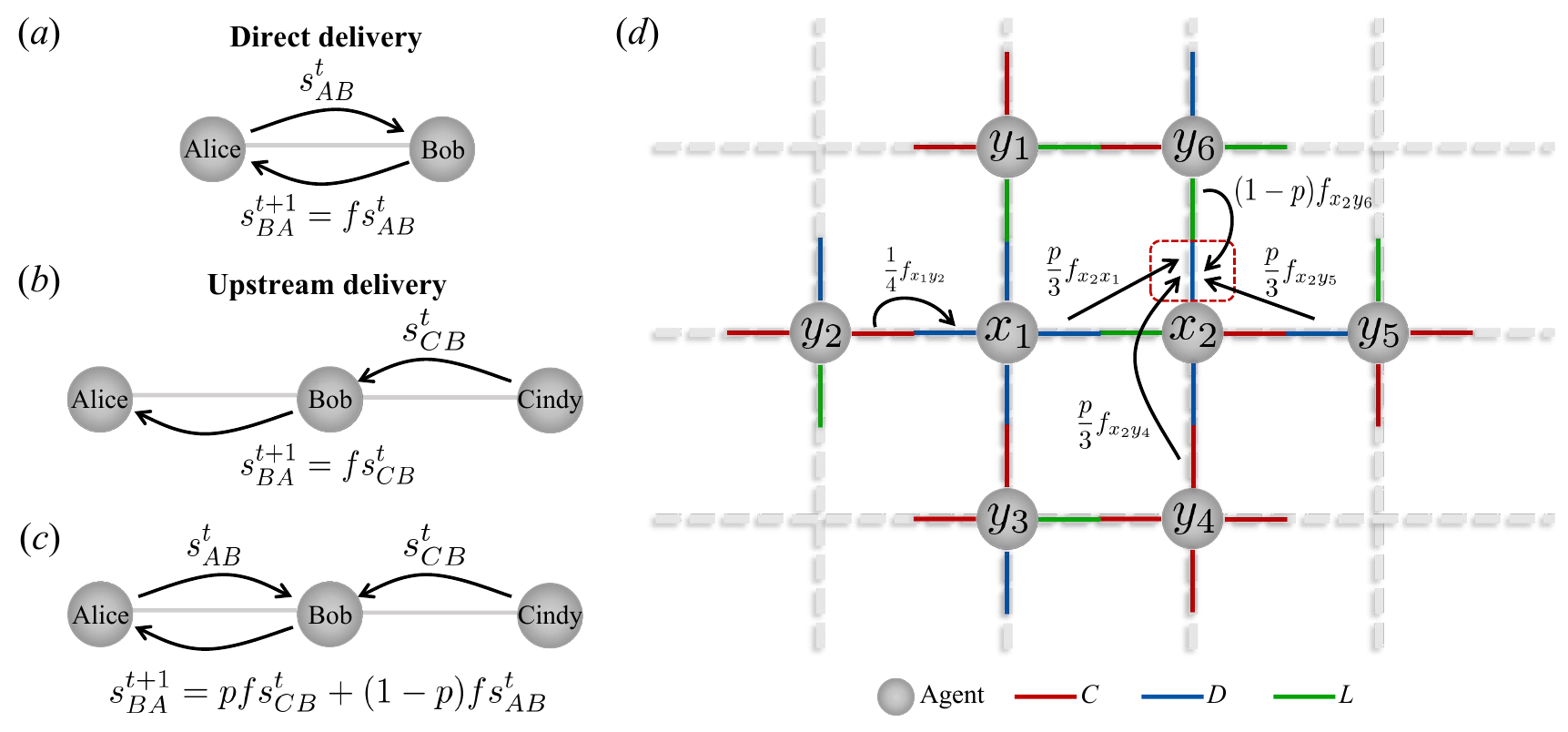}
  \renewcommand{\thefigure}{1}
\caption{Diagram of strategy delivery path and interactive diversity. Here $s_{AB}$ is noted as the player Alice’s strategy towards Bob, and so on. (a) Direct delivery means that player Bob's decision on Alice depends on Alice's performance on Bob with a certain probability, reflecting the transfer of strategy between two individuals who directly interact. (b) Upstream delivery means that Cindy's performance against Bob affects Bob's decision against Alice with a certain probability, reflecting the transfer of strategy between unrelated individuals. (c) Player Bob adjusts its strategy against Alice by combining the two delivery paths, following upstream delivery with probability $p$ and direct delivery with probability $1-p$. (d) Two types of population are distributed in the network environment, namely node players (i.e. player $x_1$, without interactive diversity) and link players (i.e. player $x_2$, with interactive diversity). Among them, the node player randomly selects a neighbor to imitate. The link player updates his strategy towards a neighbor (i.e. neighbor $y_6$), here he selects other neighbors (unrelated to the neighbor $y_6$, like neighbors $y_5$, $y_4$, and $x_1$) with probability $p$, and selects the neighbor $y_6$ with probability $1-p$. Then the link player imitate the selected neighbor's strategy towards himself with probability $f$ measured by the Fermi function. It is worth noting that link players update their strategies for each neighbor in the same way in a clockwise direction. From a macroscopic perspective, the behavior of link players updating their strategies can be simply viewed as preferring upstream delivery with probability $p$ and direct delivery with probability $1-p$.}
\label{fig1}       
\end{figure*}

\section{Model}
Based on voluntary prisoner's dilemma game, we explore how interactive diversity affects cooperative behavior on the spatial network. Individuals are placed on each node of the network. The population on the network can be divided into node group and link group depending on their decision patterns. The density of link groups is $u$, and the two groups are randomly distributed on the network. Node individuals without interactive diversity can only adopt the same strategies to interact with their neighbours simultaneously, as player $x_1$ in Fig.\ref{fig1}. Link individuals with interactive diversity are free to adopt different strategies to interact with their neighbours, as player $x_2$ in Fig.\ref{fig1}. Initially, all individuals are randomly assigned to cooperation ($C$), defection ($D$), and loner ($L$). For any player $i$ with $k$ neighbors, his strategy $s_i$ can be expressed as $s_i=(s_{i1},s_{i2},\cdots,s_{ik})$, where $s_{ik}$ ($s_{ik}\in \left\{C,D,L\right\}$) represents the player $i$'s strategy towards neighbor $k$. In particular, for node player $i$, $s_{i1}=s_{i2}=\cdots=s_{ik}$, while for link player $i$, $s_{i1}$, $s_{i2}$, and $s_{ik}$ are independent of each other.

All players interact with their neighbors and obtain cumulative payoffs according to the payoff matrix shown below~\cite{szabo2002evolutionary}:

\begin{equation} 
  \bordermatrix{
      & C & D & L \cr
    C & 1 & 0 & \sigma \cr
    D & b & 0 & \sigma \cr
    L & \sigma & \sigma & \sigma 
 },
\end{equation}
where $R=1$, $P=S=0$, and $T=b$ $(1<b\leq 2)$, and the parameter $\sigma$ represents the payoff of loner and any of his opponents, without loss of generality, and following the previous researches, we fixed $\sigma=0.3$~\cite{szabo2002evolutionary,szabo2004cooperation}.

Subsequently, any individual $x$ updates his strategies by imitating a neighbour's strategy with probability $1-\epsilon$, where the mutation factor is taken into account, i.e., the individual chooses a strategy randomly with probability $\epsilon$. Further, for imitation updates, the player imitates a chosen neighbor $y$'s strategy with probability $f$ measured by the Fermi function as follows,
\begin{equation}
f(s_{x\cdot}\gets s_{yx})=\frac{1}{1+exp[({\Phi_x}-{\Phi_y})/K]},
\end{equation}
where $\Phi_x$ ($\Phi_y$) is player $x$'s ($y$'s) cumulative payoffs, and $K$ characterizes the noise introduced to permit irrational choices. 

Given that node players and link players follow different imitation dynamics due to their different strategy forms, where node players' strategies are passed along nodes, while link players' strategies are passed along edges. Specifically, for any focal node player (i.e. player $x_1$), he randomly chooses a neighbour (i.e. player $y_2$) as reference player, and imitates the selected neighbour $y_2$'s strategy towards player $x_1$ ($s_{{y_2}{x_1}}$) with probability $f_{{x_1}{y_2}}$. However, for any focal link player (i.e. player $x_2$) have to update their strategies towards each neighbour independently, $\Omega=(y_1,y_2,\cdots,y_k)$ is denoted as the set of neighbors, and $k$ is the degree of player $x_2$. Taking the link player $x_2$ updating his strategy towards neighbour $y_i$ (i.e. neighbour $y_6$) as an example, at this time, player $x_2$ interacts directly with neighbor $y_i$, while the rest of neighbors $y_j \in \Omega (j\neq i)$ are not connected to neighbor $y_i$, and there is no interactions between them. Therefore, link players' behavior may be influenced by various neighbors and trigger different emotions, namely direct and indirect emotions, which drive the strategy along the edge in two delivery paths respectively. In more detail, direct emotion is expressed as ``Bob’s decision to Alice depends with a certain probability on Alice’s performance to Bob", driving the strategy to be passed between two people, denoted as direct delivery, as Fig.\ref{fig1}($a$). Indirect emotion is expressed as ``Cindy's behavior towards Bob affects Bob's decision towards Alice with a certain probability", driving the transfer of strategies among unrelated individuals, recorded as upstream delivery, as Fig.\ref{fig1}($b$). Here, we introduce the preference for upstream delivery $p (p\in [0,1])$ reflects the attitude of link players on which method to update their strategies, as Fig.\ref{fig1}($c$). In other words, if a link player updates his strategy towards neighbour $y_i$, the neighbor $y_{j (j=i)}$ is selected with probability $1-p$ for direct delivery, and the rest of neighbors $y_{j (j\neq i)}$ are selected with probability $p$ for upstream delivery.

Furthermore, the player $x_2$ imitates the selected neighbour $y_j$'s strategy towards player $x_2$ ($s_{{y_j}{x_2}}$), i.e., replaces strategy $s_{{x_2}{y_i}}$ with strategy $s_{{y_j}{x_2}}$ with probability $f_{{x_2}{y_j}}$.
Therefore, the probability $W$ that any player $x$ imitates a neighbor $y_j$'s strategy $s_{y_jx}$ can be uniformly described as follows,

\begin{equation}
W(s_{xy_i}\gets s_{y_jx})=\left\{
\begin{array}{cc}
(1-p){\cdot}\frac{1}{1+exp[({\Phi_x}-{\Phi_{y_j}})/K]} & {j=i}\\
\frac{p}{k-1}{\cdot}\frac{1}{1+exp[({\Phi_x}-{\Phi_{y_j}})/K]} & {j\neq i} 
\end{array}. \right.
\end{equation}

Obviously, for link players, $p=0$ means that they update their strategies towards neighbour $y_i$ only by imitating neighbor $y_i$, regardless of other neighbors. Conversely, $p=1$ means that link players update their strategies towards neighbour $y_i$ only by imitating other neighbors, not $y_i$. In addition, Eq.(2.3) applies to node players at $p$ fixed as $1-\frac{1}{k}$, in which each neighbor is selected with equal probability.

We explore how interactive diversity and preference for upstream delivery affect the evolution and cyclic dominance of strategies in three structural populations based on a voluntary prisoner's dilemma game, including regular lattice with periodical boundary conditions, regular small world network, and regular random network. Sizes of populations are $N=300\times300$ for lattice network, $N=9\times10^4$ for other networks with $k=4$. Here, we follow the Monte Carlo asynchronous update rule, thus each individual is selected on average once to update its strategy towards its neighbors in a full Monte Carlo step. All results in the steady state are the average of the last $10^4$ from the total $5\times10^4$ steps. In order to ensure the accuracy of final equilibrium results, all results are averaged over up to 10 independent realizations.

\begin{figure*}
\centering
  \includegraphics[scale=0.52]{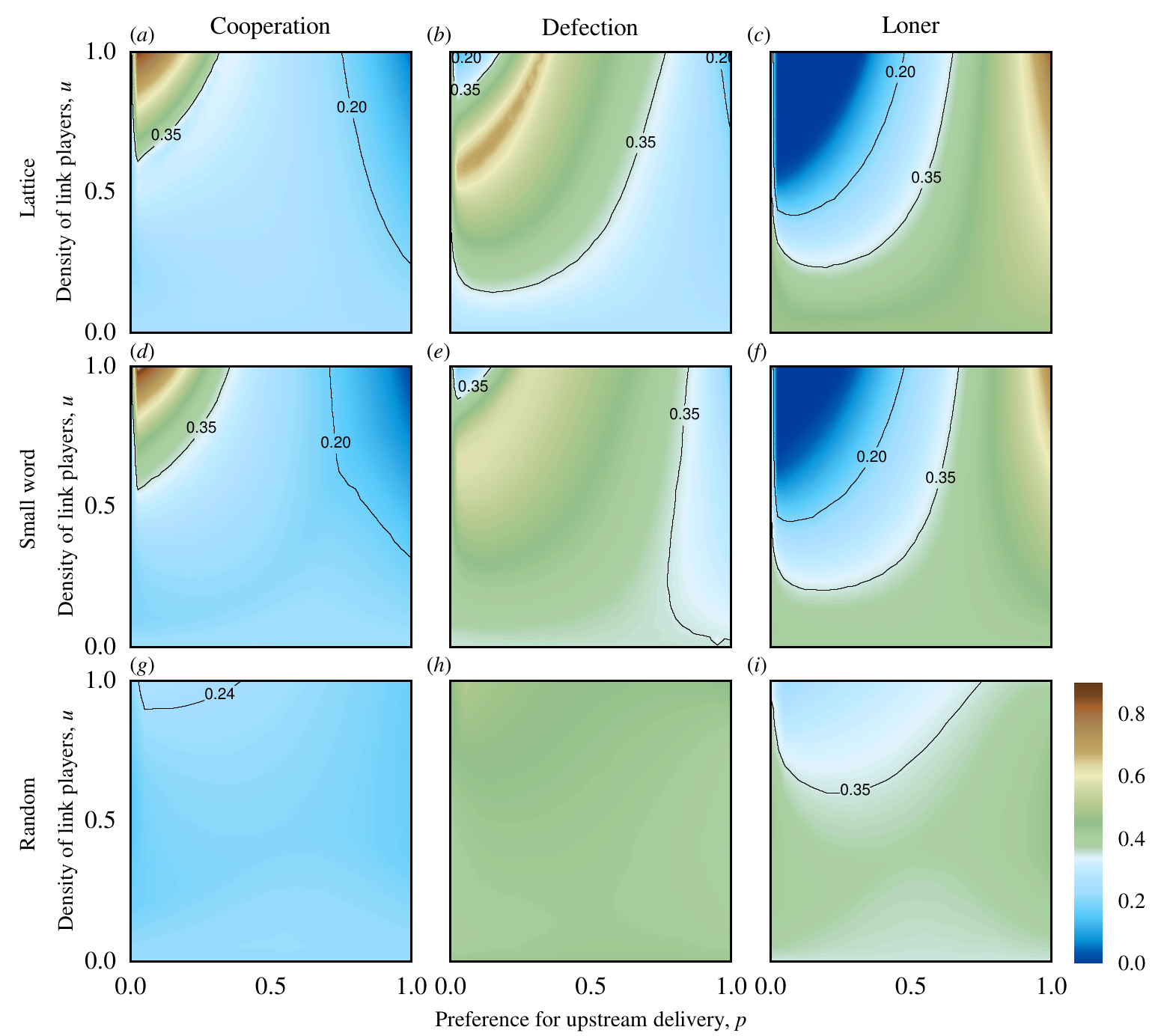}
  \renewcommand{\thefigure}{2}
\caption{Interactive diversity promotes cooperative behavior under weak preferences for upstream delivery and inhibits cooperative behavior under strong preferences for upstream delivery. Contour plots of the frequency of cooperation, defection and loner in the steady state in $u-p$ plane. All results are obtained for
$b=1.7$, and $\epsilon=10^{-6}$.}
\label{fig2}       
\end{figure*}
 
\section{Results}
First we explore how interactive diversity as measured by the density of link players $u$ and preference for upstream delivery $p$ affects group behavior. Fig.\ref{fig2} illustrates the change of strategies with parameters $u$ and $p$ in the evolutionary steady state. It is found that the effect of interactive diversity on promoting cooperation depends on the strategy update method of link players, that is, interactive diversity favors cooperative behavior in the case of weak preference for upstream delivery, yet the opposite is true for strong preference for upstream delivery. In addition, the proportion of strategy $L$ and strategy $C$ show the opposite trend. Further, Fig.\ref{A1} explores the frequency of strategies with $u$ and temptation $b$ under different preferences for upstream delivery. On the one hand, it verifies the above findings that compared with $p=0$, interactive diversity improves cooperative behavior at $p=0$. On the other hand, the group maintains high cooperation under small $b$.

\begin{figure*}
\centering
  \includegraphics[scale=0.5]{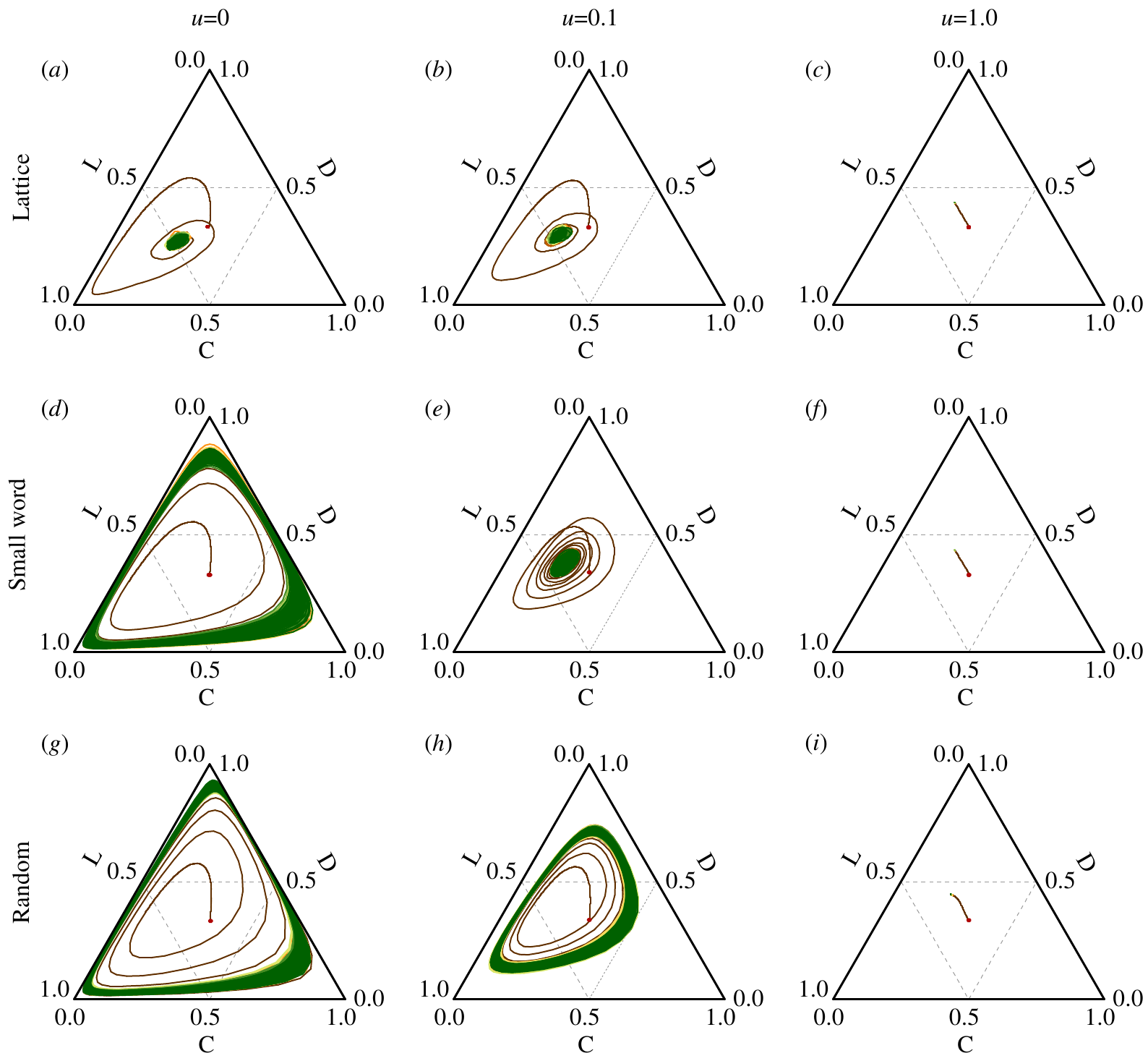}
  \renewcommand{\thefigure}{3}
\caption{The evolution of cyclic interactions goes from global cyclic dominance (i.e., persistent periodic oscillations of three strategies) to local cyclic dominance (i.e., a stable stationary state with all three strategies coexisting) until it disappears as interactive diversity increases. The evolutionary trajectories of strategies in different networks for different densities of link players $u$. A red dot marks the random initial conﬁguration with the same proportion of three strategies. Persistent periodic oscillations do not appear on the lattice network, while 3 evolutionary states are observed on the small world and random networks. Here we take $p=0$, $b=1.7$, $\epsilon=10^{-6}$, and the $u$ values are 0, 0.1, and 1.0, respectively.}
\label{fig3}       
\end{figure*} 

\begin{figure*}
\centering
  \includegraphics[scale=0.55]{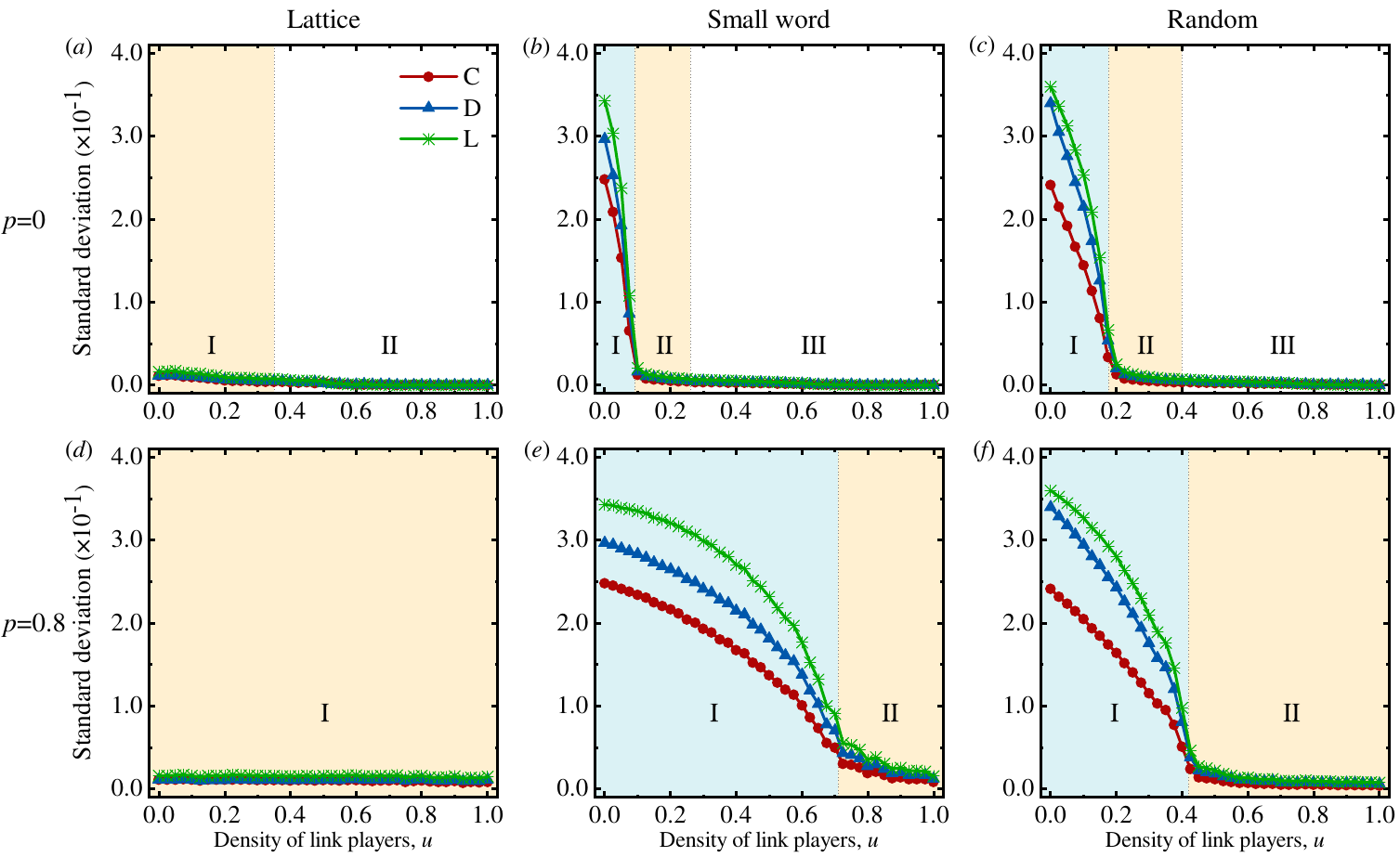}
  \renewcommand{\thefigure}{4}
\caption{Interactive diversity tames the cyclic dominance, regardless of the link player's preference for upstream delivery. The standard deviation of strategies in the evolutionary stable state as a function of the density of link players $u$. The standard deviation of strategies is utilised to approximately identify the evolutionary trend of system, and the evolution presents 3 scenarios: global cyclic dominance (blue), local cyclic dominance (yellow) and without cyclic dominance (white). Here we take $p=0$ (top) and $p=0.8$ (bottom), $b=1.7$, and $\epsilon=10^{-6}$.}
\label{fig4}       
\end{figure*} 

\begin{figure*}
\centering
  \includegraphics[scale=0.68]{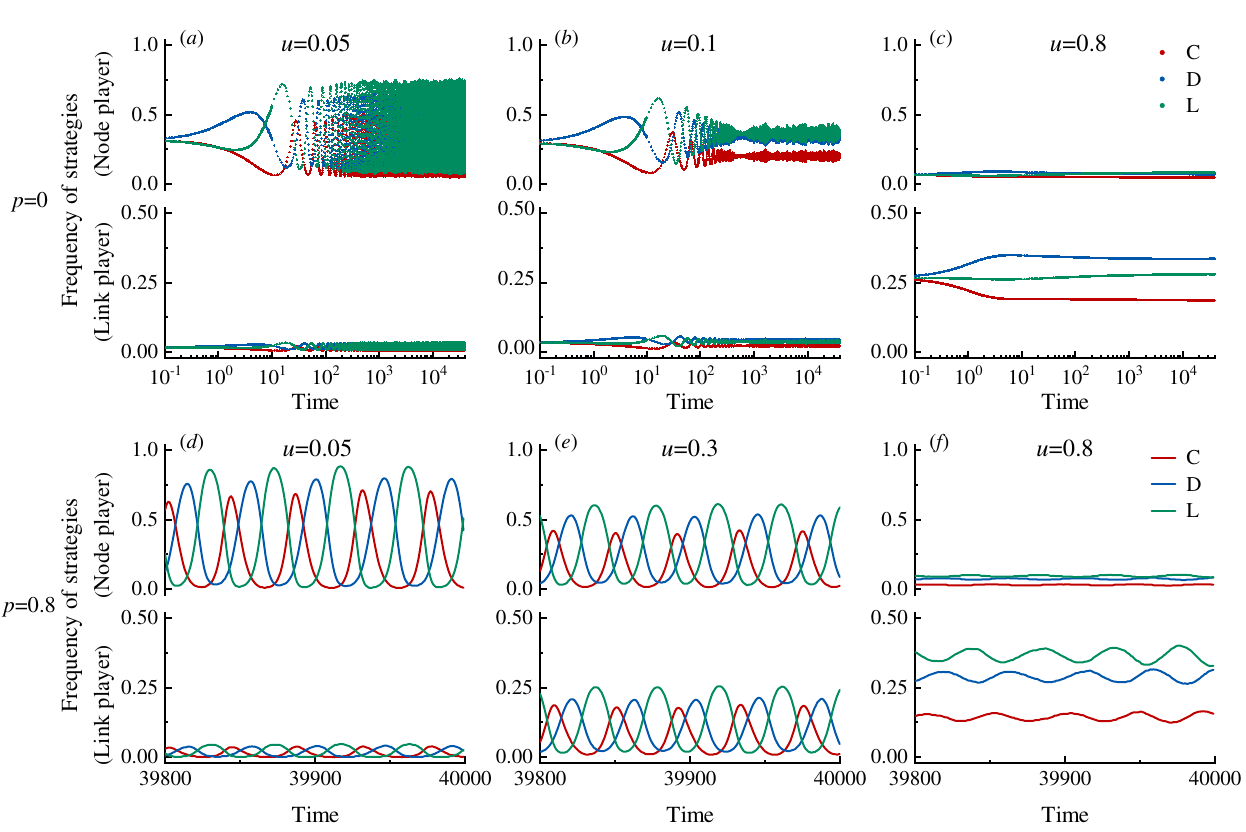}
  \renewcommand{\thefigure}{5}
\caption{Interactive diversity weakens the fluctuation of cyclic interactions for both node and link players. The evolution of strategies for node and link players in mixed populations on the small world network for different densities of link players $u$. Here we take $p=0$ (top), $p=0.8$ (bottom), $b=1.7$, and $\epsilon=10^{-6}$.}
\label{fig5}       
\end{figure*} 

Next we explore the evolution of strategies in different structural populations, Fig.\ref{fig3} and Fig.\ref{A2} show the dynamic evolutionary trajectories of three strategies $C$, $D$, and $L$ under different interactive diversity $u$. The former ($p=0$) describes the case where link players update their strategies in a direct delivery manner, and the latter reflects link players’ strong preference for upstream delivery ($p=0.8$). The results show that interactive diversity significantly weakens the phenomenon of cyclic dominance. Specifically, when the population is full of node players ($u=0$, without interactive diversity), the dynamic trajectory starts from the center and gradually forms cyclic dominance. In the lattice network, after a period of cyclic dominance, the system evolves towards a stable stationary state with all three strategies coexisting, denoted as local cyclic dominance (Fig.\ref{fig3}(a)). While in the small world and random networks, the system evolves towards a state with persistent periodic oscillations of three strategies, denoted as global cyclic dominance (Fig.\ref{fig3}(d,g)). Increasing the interactive diversity of the population leads to the evolution of strategies from global cyclic dominance to local cyclic dominance (Fig.\ref{fig3}(e)), and even the cyclic dominance disappears at $u=1$. Surprisingly, it is found that when link players' update methods are more inclined to upstream delivery (i.e. $p=0.8$), global cyclic dominance still occurs in small world and random networks in interaction environments with a higher density of link players (i.e. $u=0.3$). Not only that, the strong preference for upstream delivery effectively prevents the cyclic dominance from disappearing in situation with the greatest interactive diversity, as shown in Fig.\ref{A2}.

Subsequently, to quantitatively describe how interactive diversity affects cyclic dominance, we take the standard deviation of strategies in the evolutionary stable state as the condition to approximately identify the evolutionary trend of system (Fig.\ref{fig4}). The results show that the oscillation amplitude of strategy evolution decreases with interactive diversity, regardless of link players’ strategy update methods and network topologies. Here we give two critical values $u_{gl}$ (about $0.5\times 10^{-1}$) and $u_{ld}$ (about $0.5\times 10^{-2}$) reflecting the evolutionary state from global cyclic dominance to local cyclic dominance, and from local cyclic dominance to disappearance, respectively. Approximately, in the case of $p=0$, $u_{ld}=0.35$ in the lattice network (Fig.\ref{fig4}(a)), $u_{gl}=0.09$ and $u_{ld}=0.26$ in the small world network (Fig.\ref{fig4}(b)), etc. Further in the case where link players’ strong preference for upstream delivery (i.e. $p=0.8$), the standard deviation of strategies increases significantly, which not only avoids the disappearance of cyclic dominance in various networks, but also increases the critical value $u_{gl}$, so that the global cyclic dominance appears in a wider range of interactive diversity.

In order to further explore how interactive diversity affects the strategy evolution of the two types of players. We show the evolution of strategies for node and link players in mixed populations respectively, taking the small world network as an example (Fig.\ref{fig5}). It is found that in the case of direct delivery manner ($p=0$), node players dominate the cycle dominance of strategies. Although the local cyclic dominance of link players can be observed at the beginning of evolution, its effect is negligible (Fig.\ref{fig5}(a,b)). With the increase of interactive diversity, the cyclic dominance of node players is rapidly weakened until it disappears, and the system quickly reaches an evolutionary stable state (Fig.\ref{fig5}(c)). In addition, the introduction of strong preference for upstream delivery ($p=0.8$) reveals that the link population is sufficient to show clear persistent periodic oscillations even its density is very small. Moreover, as the interactive diversity increases both types of players still show global cyclic dominance (i.e. $u=0.3$). In particular, the local cyclic dominance of link group is observed at $u=0.8$. Obviously the strategy evolution of the link group becomes the key to dominate the evolutionary trend of the whole group. Implying that upstream delivery works by stimulating the cyclic dominance of link players, which further reinforces the fluctuation of cyclic interactions of node players, so as to avoid the disappearance of cyclic dominance.

\begin{figure*}
\centering
  \includegraphics[scale=0.6]{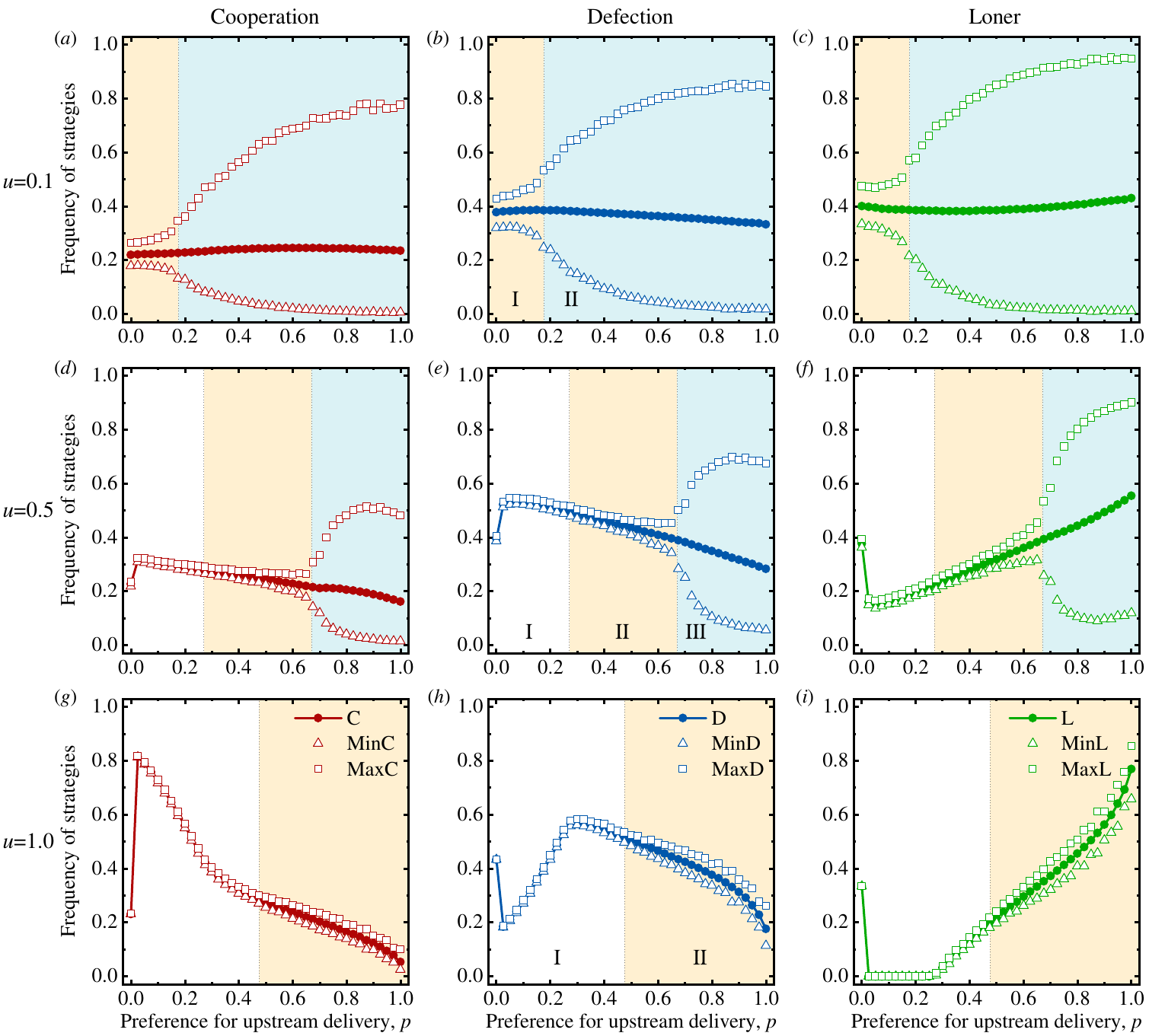}
  \renewcommand{\thefigure}{6}
\caption{Link players' preference for upstream delivery increases the fluctuation of strategies, but still fails to achieve global cyclic dominance in environments with high interactive diversity. The average frequency of strategies ($\bullet$), the minimum ($\triangle$) and maximum ($\square$) values of strategies as a function of preference for upstream reciprocity under different densities of link players. From left to right, it describes cooperation (red), defection (blue), and loner (green) strategy. The persistent periodic oscillations are indicated by the minimum and maximum values of strategies in the blue area. Here the results are obtained on the small world network, and $b=1.7$, $\epsilon=10^{-6}$. The density of link player $u$ is equal to 0.1, 0.5, and 1.0, respectively.}
\label{fig6}       
\end{figure*}

Subsequently, in order to clearly show the effect of preference for upstream reciprocity on the fluctuations of the strategies, we provide the average frequency of strategies ($\bullet$), the minimum ($\triangle$) and maximum ($\square$) values of strategies on the small world network (Fig.\ref{fig6}). Here we show another way to judge the evolutionary state of system by comparing the deviation between the minimum and maximum values, and the average frequency of strategies in the evolutionary stable state. It is found that the fluctuation of strategies gradually increases with link players’ preference for upstream reciprocity, which is reflected in that the minimum and maximum values of strategies go from almost overlapping to slightly separated and finally showing a large bifurcation. Further, the spacing between the minimum and maximum values of strategies indicates the evolution of system presents three scenarios, namely, without cyclic dominance (white area), local cyclic dominance (yellow area), and global cyclic dominance (blue area). In addition, with the increase of interactive diversity, the effect of strong preference for upstream delivery is no longer significant, especially in the case of $u=1$, the global cyclic dominance disappears. These results again clearly confirm that interactive diversity reduces the amplitude of strategy oscillations. 

\begin{figure*}
\centering
  \includegraphics[scale=0.6]{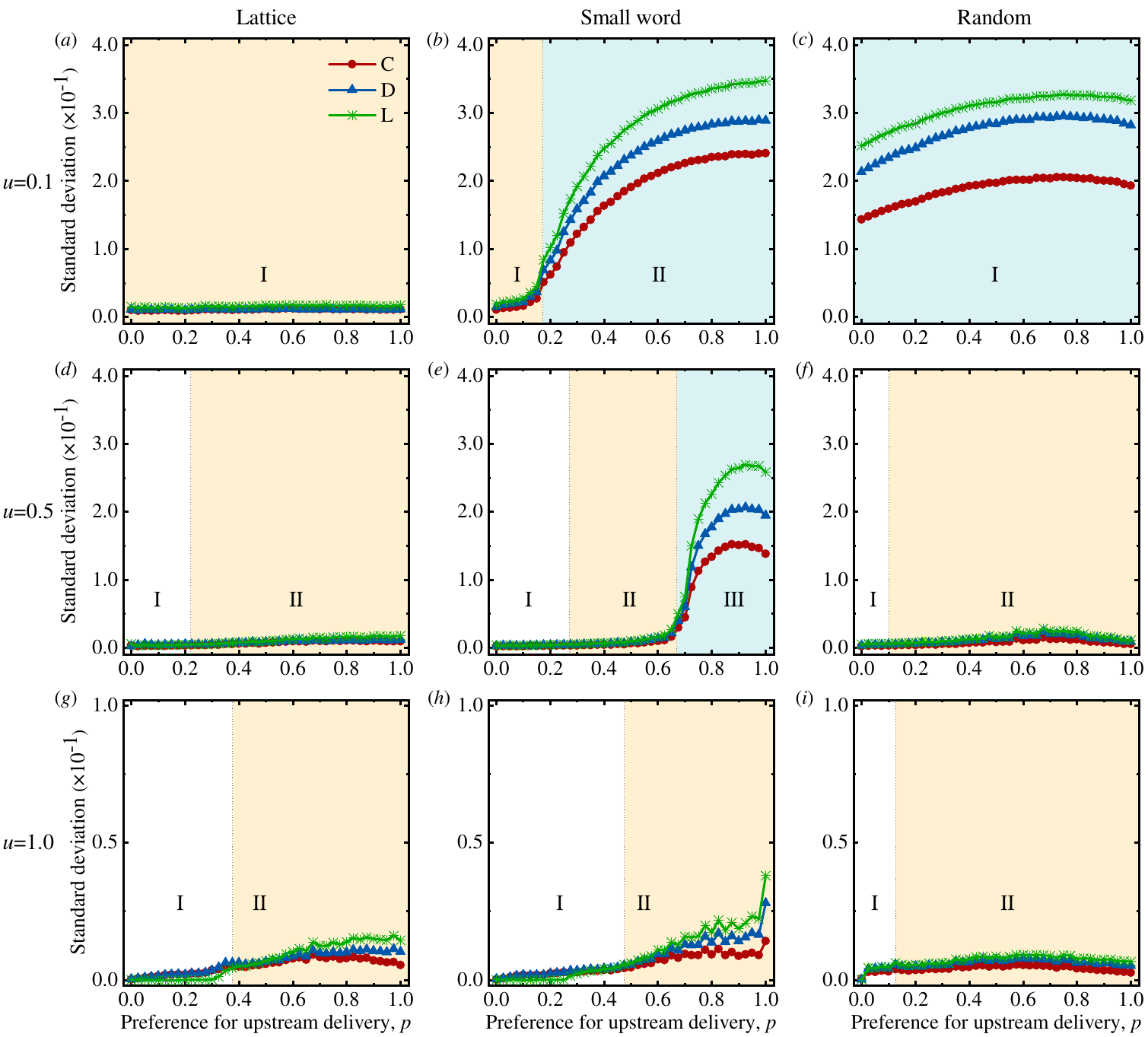}
  \renewcommand{\thefigure}{7}
\caption{The standard deviation of strategies in the evolutionary stable state as a function of the preference for upstream reciprocity $p$ in different networks for different densities of link players. Here we take $b=1.7$, $\epsilon=10^{-6}$. The density of link player $u$ is equal to 0.1, 0.5, and 1.0, respectively.}
\label{fig7}       
\end{figure*} 

On the basis of the above results, we accurately calculated how the standard deviation of strategies varies with the preference for upstream delivery under different networks (Fig.\ref{fig7}),  and combine the results of Fig.\ref{fig4} and Fig.\ref{A3}, two different forms of results together confirm the effect of interactive diversity and preference for upstream delivery on cyclic dominance.

Finally, we analyze the effect of temptation $b$ on strategy evolution and cyclic dominance. It is found that temptation $b$ and upstream delivery have similar effects, driving system evolution toward a state where the three strategies coexist (Fig.\ref{A4}). As shown in Fig.\ref{A5}, the frequency of cooperation monotonically decreases with temptation $b$, while loner monotonically increases. In particular, loner disappears for small $b$. Since loner invades the defectors, they can only survive if the frequency of defection is sufficiently high, thus an increase in temptation $b$ leads the system to move from two-strategy to three-strategy coexistence. As previously described, the extent to which the minimum and maximum values of cooperation deviate from the average, as well as the standard deviation of strategies, reflect the state of cyclic dominance. The simulation results again confirm that the amplitude and periodicity of cyclic dominance increases with $b$. In addition, cyclic dominance produces self-organizing patterns on lattice networks, whereas it exhibits different oscillatory behavior, namely persistent periodic oscillations, on small world and random networks.

\section{Conclusion and discussion}
In this paper, building on the spatial voluntary prisoner's dilemma game, we have taken into account interactive diversity to investigate its impact on cyclic dominance and the evolution of cooperation. Specifically, the notion of interactive diversity is manifest in link players. These players adjust their actions towards their peers based on two distinct emotional attitudes, demonstrating the flexibility of their decision-making behavior. For instance, player Bob’s behavior towards Alice is not only dependent on Alice’s strategy towards Bob, but also influenced by another player Cindy (unrelated to Alice) towards Bob. The former follows the attitude of direct delivery, while the latter follows the attitude of upstream delivery, i.e., player Bob may pass on positive or negative behavior (emotions) from player Cindy to player Alice. In contrast with 
the formation of cyclic dominance effect that sustains cooperation under node dynamics scenario~\cite{szabo2002evolutionary,hauert2005game}. The results show that introducing  interactive diversity leads to non-trivial results, when link players tend to update their strategies with direct delivery, the interactive diversity not only hinders the dynamic cyclic dominance of three strategies, but also weakens the amplitude of periodic oscillations, and the system evolves from global cycle dominance to local cycle dominance until it disappears. Interestingly, when link players prefer upstream delivery, cyclic interactions is stimulated, prompting persistent periodic oscillations appear in a wider range of interactive diversity, and even reversing the disappearance of cyclic dominance in networks with greater interactive diversity.

It's essential to note that our results are based on structured populations. We intentionally excluded well-mixed populations from our study, given that the voluntary prisoner's dilemma fails to display a cyclic dominance phenomenon in such populations. Therefore, exploring the impact of interactive diversity in such settings would be irrelevant. However, it should be noted that cyclic dominance does occur in both well-mixed and networked populations in the voluntary public goods game~\cite{hauert2002replicator,liu2022early,szabo2002phase,sasaki2007probabilistic}. Thus, future work could profitably focus on incorporating interactive diversity into the voluntary public goods game to examine its impact on cyclic dominance. Moreover, moving beyond the basic homogeneous network structure and considering more realistic network structures, such as higher-order~\cite{laraki2013higher,grilli2017higher} or temporal networks~\cite{roca2009evolutionary,li2020evolution}, could provide valuable insights. It would be interesting to see how such extensions impact the role of cyclic dominance in fostering cooperation and if cyclic dominance remains the primary force underpinning the persistence of cooperative behaviors.  

Existing studies on cyclic dominance in structured populations have underscored its importance in sustaining cooperation in various social dilemmas. These dilemmas include volunteer prisoner's dilemma~\cite{szabo2002evolutionary}, public goods game with peer/poor punishment~\cite{szolnoki2017second,szolnoki2011phase}, spatial ultimatum game with discrete strategies\cite{szolnoki2012defense}, and among others~\cite{szolnoki2004phase}. However, these studies often rely on an unrealistic assumption: that agents interact uniformly with their neighbors. Our research relaxes this assumption by granting agents the flexibility to choose different strategies when interacting with different neighbors. In doing so, we challenge the prevailing notion that cyclic dominance is essential for the maintenance of cooperation. Our findings suggest that when interactive diversity is incorporated into the model, the establishment of cyclic dominance is hindered. Importantly, cooperation is not eliminated; rather, its persistence becomes less dependent on the mechanisms of cyclic dominance. While our analysis has specifically addressed the influence of interactive diversity on the cyclic dominance of strategies in the spatial voluntary prisoner's dilemma, we believe that interactive diversity might similarly attenuate the role of cyclic dominance in other spatial evolutionary contexts. This attenuation could be compensated for through alternative mechanisms related to interactive diversity, thereby sustaining cooperation in a different manner. In light of the ubiquity of interactive diversity, our findings suggest that the importance of cyclic dominance in the evolutionary emergence of cooperation may have been previously overestimated.

\textbf{Ethics}: This work did not require ethical approval from a human subject or animal welfare committee.

\textbf{Data Accessibility}: \url{https://github.com/Danyang061226/CDL.git}

\textbf{Authors’ Contributions}: D.J. and C.S. conceived the research. D.J. performed simulations. D.J. and C.S. wrote the manuscript. All coauthors discussed and interpreted the results, and edited the manuscript.

\textbf{Competing Interests}: We declare we have no competing interests.

\textbf{Funding}: This research was supported by the National Science Fund for Distinguished Young Scholars (No. 62025602), the National Science Fund for Excellent Young Scholars (No. 62222606), the National Natural Science Foundation of China (Nos. 11931015, U1803263, 81961138010 and 62076238), Fok Ying-Tong Education Foundation, China (No. 171105), Technological Innovation Team of Shaanxi Province (No. 2020TD-013), Fundamental Research Funds for the Central Universities (No. D5000211001), the Tencent Foundation and XPLORER PRIZE, JSPS Postdoctoral Fellowship Program for Foreign Researchers (grant no. P21374).

\bibliography{refs}

\clearpage

\begin{figure*}
\centering
  \includegraphics[scale=0.36]{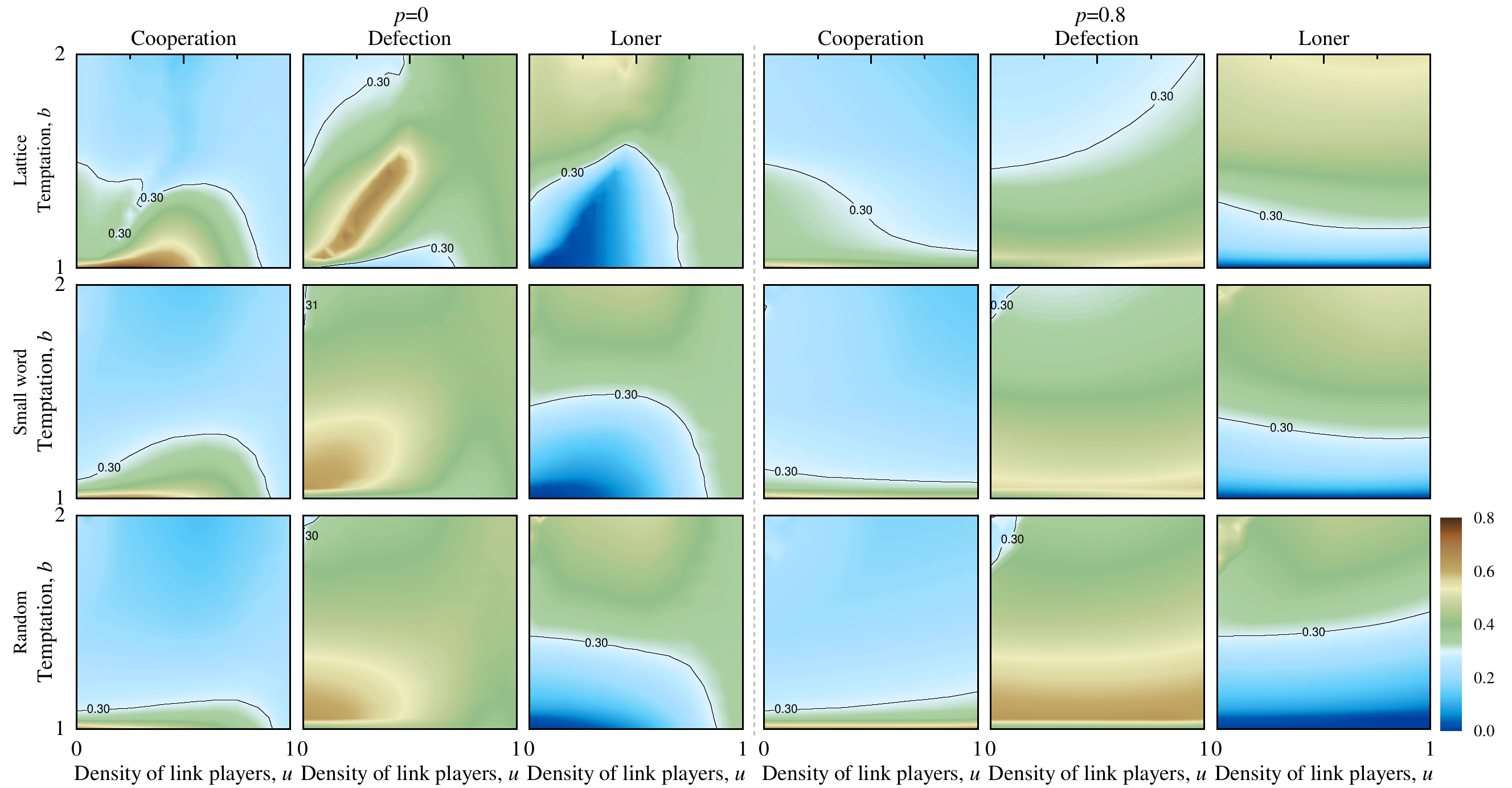}
  \renewcommand{\thefigure}{A1}
\caption{Different from the case with a strong preference for direct delivery, interactive diversity significantly inhibits cooperative behavior under the strong preference for upstream delivery. Contour plots of the frequency of cooperation, defection and loner in the steady state in $u-b$ plane. Here we take $p=0$ (left) and $p=0.8$ (right), $\epsilon=10^{-6}$.}
\label{A1}       
\end{figure*}

\begin{figure*}
\centering
  \includegraphics[scale=0.5]{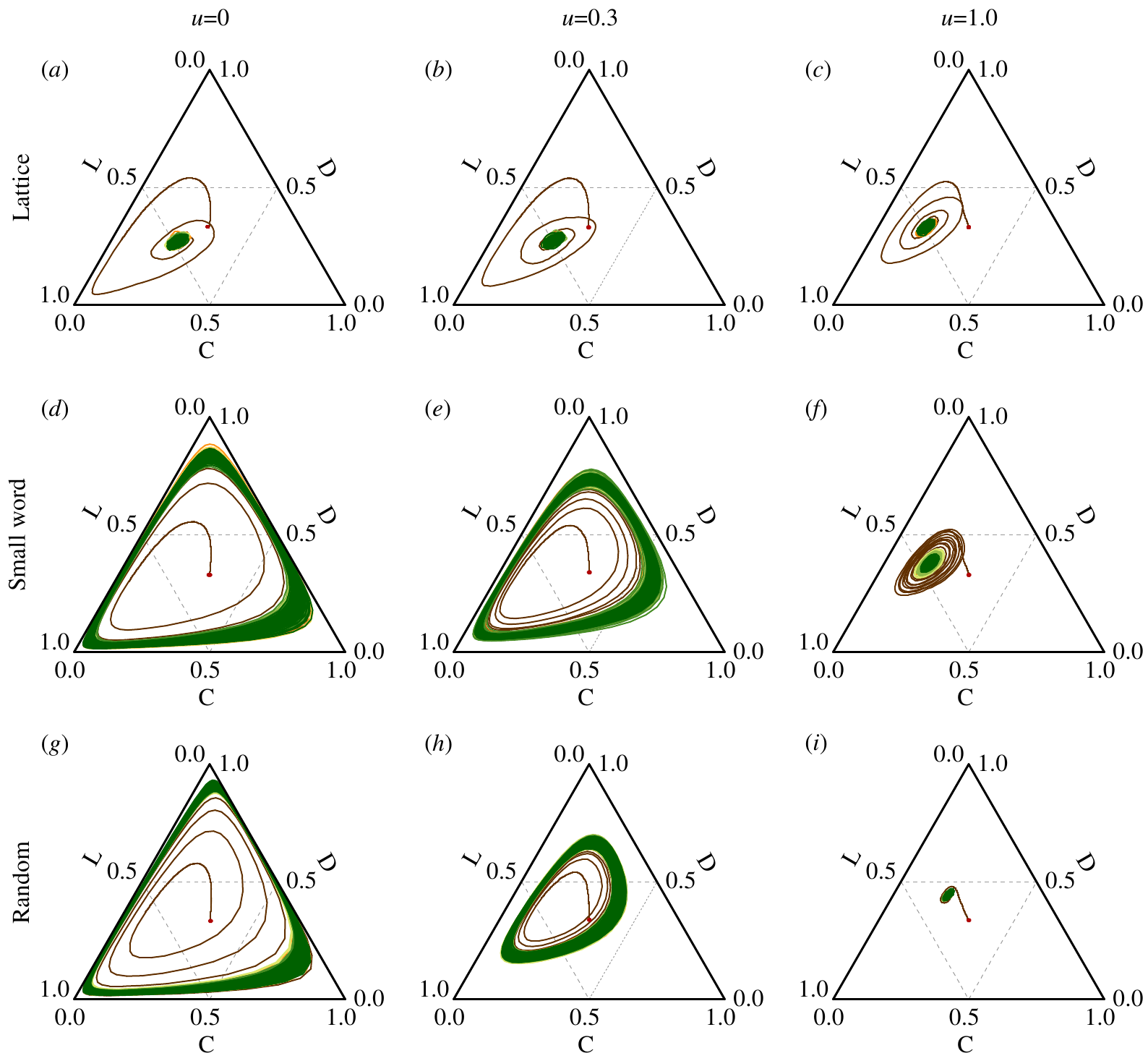}
  \renewcommand{\thefigure}{A2}
\caption{The strong preference of link players for upstream delivery effectively prevents cyclic dominance from disappearing. The evolutionary trajectories of strategies in different networks for different densities of link players $u$. A red dot marks the random initial conﬁguration with the same proportion of three strategies. Local cyclic dominance occurs in all three networks even the game environment with the strongest interactive diversity. Here we take $p=0.8$, $b=1.7$, $\epsilon=10^{-6}$, and the $u$ values are 0, 0.3, and 1.0, respectively.}
\label{A2}       
\end{figure*}

\begin{figure*}
\centering
  \includegraphics[scale=0.6]{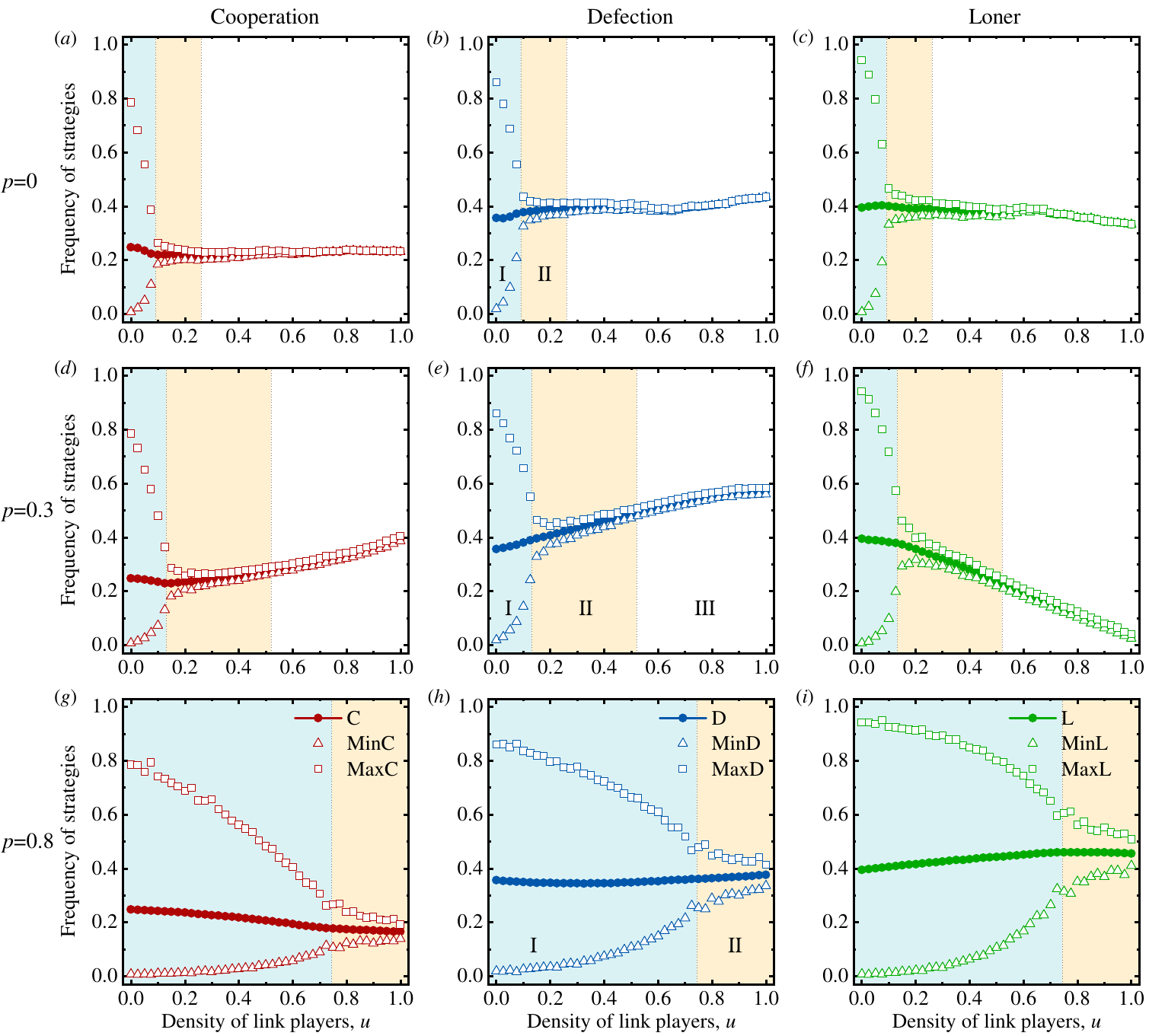}
  \renewcommand{\thefigure}{A3}
\caption{Interactive diversity weakens cyclic dominance and the amplitude of periodic oscillations, regardless of the link players' preference for upstream delivery. The average frequency of strategies ($\bullet$), the minimum ($\triangle$) and maximum ($\square$) values of strategies as a function of density of link players under different preference for upstream delivery. From left to right, it describes cooperation (red), defection (blue), and loner (green) strategy. The persistent periodic oscillations are indicated by the minimum and maximum values of strategies in the blue area. Here the results are obtained on the small world network, and $b=1.7$, $\epsilon=10^{-6}$. The preference for upstream delivery $p$ is equal to 0, 0.3, and 1.0, respectively.}
\label{A3}       
\end{figure*}

\begin{figure*}
\centering
  \includegraphics[scale=0.55]{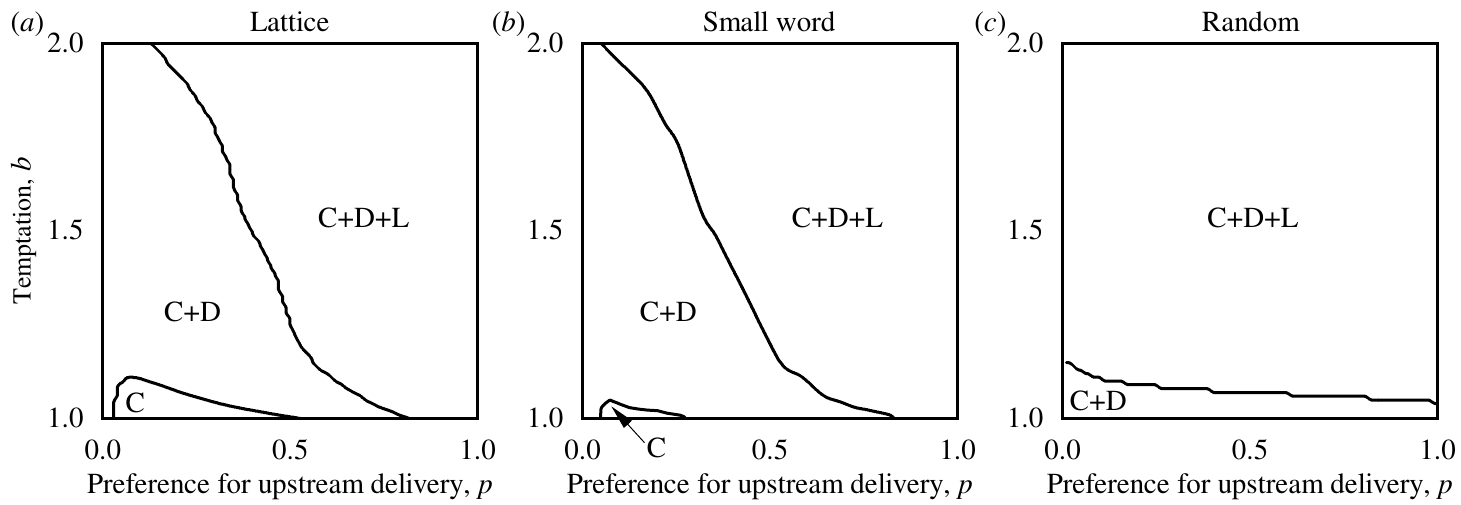}
  \renewcommand{\thefigure}{A4}
\caption{The strong preference for upstream delivery and the temptation to defect favor the evolution of system from single homogeneous state to three-strategy coexistence state. Strategy combinations on the network in the steady state in $b-p$ plane. All results are obtained for $u=1.0$, and $\epsilon=0$.}
\label{A4}       
\end{figure*}

\begin{figure*}
\centering
  \includegraphics[scale=0.6]{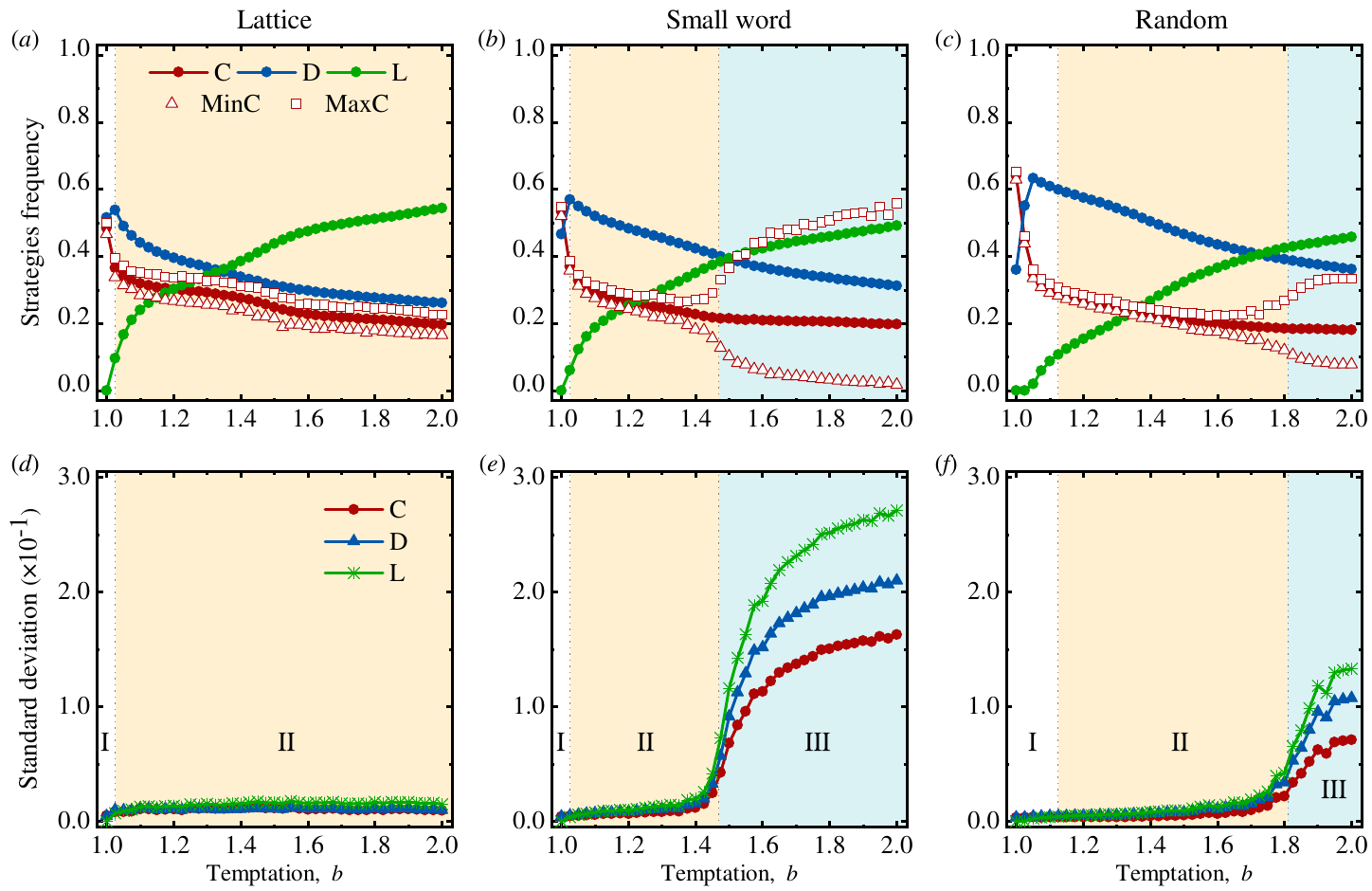}
  \renewcommand{\thefigure}{A5}
\caption{The temptation to defect reinforces the fluctuation of cyclic interactions, and the evolution of system towards a state of persistent periodic oscillations. The average frequency of three strategies ($\bullet$), the minimum ($\triangle$) and maximum ($\square$) values of cooperation strategy in the evolutionary stable state as a function of the temptation to defect $b$, for different networks. The persistent periodic oscillations are indicated by the minimum and maximum values of cooperation strategy in the blue area. Here we take $u=0.5$, $p=0.8$, and $\epsilon=10^{-6}$.}
\label{A5}       
\end{figure*}

\end{document}